\documentstyle[11pt]{article}
\begin{document}
{\setlength{\oddsidemargin}{1.2in}
\setlength{\evensidemargin}{1.2in} } \baselineskip 0.55cm
\begin{center}
{ {\bf\textbf{Quantum radiation properties of general non-stationary black hole}}}
\end{center}
\begin{center}
${\rm T.\ Ibungochouba\ Singh}$
\end{center}
\begin{center}
 Department of Mathematics, Manipur University,\\ Canchipur, 795003 Manipur (India)\\ Email: ibungochouba@rediffmail.com
\end{center}
\date{}

\begin{abstract}
 Using the generalized tortoise coordinate transformations the quantum radiation properties of Klein-Gordon scalar particles, Maxwell's electromagnetic field equations and Dirac equations are investigated in general non-stationary black hole. The locations of the event horizon and the Hawking temperature depend on both time and angles. A new extra coupling effect is observed in the thermal radiation spectrum of Maxwell's equations and Dirac equations which is absent in the thermal radiation spectrum of scalar particles. We also observe that the chemical potential derived from scalar particles is equal to the highest energy of the negative energy state of the scalar particle in the non-thermal radiation in general non-stationary black hole. Applying generalized tortoise coordinate transformation a constant term $\xi$ is produced in the expression of thermal radiation in general non-stationary black hole. It indicates that generalized tortoise coordinate transformation is more accurate and reliable in the study of thermal radiation of black hole.\\

{\it Key-words}: General non-stationary black hole; Thermal radiation; Generalized tortoise coordinate transformation.\\\\
\end{abstract}

1. {\bf Introduction}

Hawking discovered the thermal radiation of black hole using the techniques of quantum field theory in curve space-time background and the derived radiation spectrum is purely thermal in nature [1, 2]. An important aspects in the study of black hole is to reveal the thermal and non-thermal radiation of black hole. Refs. [3-5] have shown that the black hole has a non zero finite entropy and the entropy of black hole is proportional to the horizon area. Refs. [6, 7] also proposed the Hawking radiation as quantum tunneling process where the particles moves in the dynamical geometry. They recovered a leading correction to the emission rate arising from the loss mass of the black hole. Following their works, Zhang and Zhao [8-10] have extended the method to more general circumstances for rotating black hole and they have shown the spectrum is no longer precisely thermal. Further, some information of the black hole can be obtained.

Akhmedov, et al. [11] investigated the Hawking radiation as tunneling picture in the Schwarzschild black hole using the relativistic Hamilton-Jacobi method. The importance of this investigations are (i) if Schwarzschild co-ordintes or any other coordinates related to them via a transformation of spatial coordinates is used, we will get twice original Hawking temperature and (ii) Any transformation involving time coordinates is utilized in the black hole evaporation, it will give the original Hawking result. Following this method, many fruitful results have been obtained in [12-14]. This factor of two issue has been resolved via the discovery of the temporal contribution to the tunneling amplitude in the literatures [15-17]. The thermodynamics of black hole in lovelock gravity and in AdS space time have been investigated by Cai [18-20]. One of the important aspects in this investigation is to calculate the Hawking temperature.

Recently,  Anghenben, et al. [21] investigated Hawking radiation as tunneling of extremal and rotating black hole using the relativistic Hamilton-Jacobi method and WKB approximation without considering the particles back reaction. Since then many other authors applied the Hamilton-Jacobi method to study the Hawking radiation for more general space time [22-25]. Choosing appropriate Gamma matrices, the Hawking radiation as tunneling from Dirac particles was investigated by Ref. [26] for the general stationary black hole.  By inserting the wave function into the Dirac field equations, the action of the radiant particles is derived. This result is related to the Boltzmann factor of emission at the Hawking radiation temperature in accordance with semiclassical WKB approximation. Refs. [27, 28] introduced tortoise coordinate transformation to study the Hawking radiation of black hole in which gravitational field is independent of time. Using tortoise coordinate transformation, Klein-Gordon equation, Maxwell's electromagnetic field equations and Dirac particles can be transformed into single form of wave equation near the event horizon. Separating the variables from the standard wave equation, ingoing wave and outgoing wave can be obtained. Extending the wave equation from outside the event horizon into the inside by rotating $-\pi$ through the lower half of the complex plane, the thermal radiation spectra can be derived. Generalizing this method, many more works have been done [29-31].

 Refs. [32-33] have also investigated the Hawking radiation by calculating the vacuum expectation values of the renormalized energy momentum tensor of spherically symmetric non-stationary black hole. This results are consistent with the Refs. [27-28]. However all these research works were confined to the quantum thermal radiation of black hole only. In addition to the quantum thermal radiation, the importance of quantum non-thermal radiation of black hole has been studied by different authors in different types of space time in the literatures [34-38].

The main aim of this paper is to investigate thermal radiation from Klein-Gordon equation, Maxwell's electromagnetic field equations, Dirac particles and also non-thermal radiation of Hamilton-Jacobi equation in general non-stationary black hole. It gives the relationship between two kind of radiation in the case of scalar particles. A new extra coupling effect is derived from the thermal radiation spectrum of Maxwell's field equations and Dirac particles. In section 2, we derive the location of horizon in general non-stationary black hole using null surface condition and generalized tortoise coordinate transformation. In section 3, adjusting the parameter $\kappa$, the Klein-Gordon equation is transformed into a wave equation near the event horizon in general non-stationary black hole. In section 4, we derive asymptotic behaviors of first order form of four equations and second order form of three equations from the Maxwell's electromagnetic field equations near the event horizon using generalized tortoise coordinate transformation (GTCT). In section 5, the asymptotic behaviors of the first order form of four equations and second order form of two equations are deduced from the Dirac particles using the GTCT near the event horizon. In section 6, the second order form of Klein-Gordon equation, the three second order form of Maxwell's equations and the two second order form of Dirac equations are transformed into single standard form of equation near the black hole event horizon. By separating the wave equation the chemical potential and the thermal radiation spectra can be obtained. In section 7, the highest energy of non-thermal radiation is obtained from the relativistic Hamilton-Jacobi equation. In section 8, we derive the expression of a new extra coupling effect which is absent from the thermal radiation spectrum of scalars particles. The relationship between thermal and non-thermal radiation in general non-stationary black hole is also established. Some conclusions are given in the last section.

2. {\bf General non-stationary black hole}. \setcounter {equation}{0}
\renewcommand {\theequation}{\arabic{equation}}
The line element describing the most general non-stationary black hole is given by
\begin{eqnarray}
ds^2&=&g_{\mu\nu}dx^{\mu}dx^{\nu}
\end{eqnarray}
where we assume the retarded Eddington-Finkelstein coordinates $(x^{0}=u, x^{1}=r, x^{2}=\theta$ and $x^{3}=\phi)$
and make the conventions that all indices of Greek letters $\mu, \nu=0,1,2,3$ and all indices of Latin letters $j, k=0,2,3$. The event horizon in general non-stationary black hole can be characterized by null hypersurface condition: $F=F(u, r, \theta, \phi)=0$. The null hypersurface condition gives the position of horizon of stationary or non-stationary black hole [39]
\begin{eqnarray}
g^{\mu\nu}\partial_\mu F\partial_\nu F=0\cdot
\end{eqnarray}
 The location and the temperature of the horizon in a non-stationary black hole may be obtained by applying tortoise coordinate transformation. By using tortoise coordinate transformation of the form $r_*=r+(2\kappa)^{-1}ln(r-r_h)$, the Klein-Gordon equation, the Maxwell's electromagnetic field equations and the Dirac particles can be combined into a standard form of wave equation in a non-stationary or stationary space-time (where $\kappa$ is the surface gravity). The location of horizon may be assumed as functions of retarded time coordinate $u=t-r_*$ and different angles $\theta, \phi$. The space time geometry outside the event horizon is described by tortoise coordinate only and in this condition, $r_*$ approaches to positive infinity when tending to infinite point and $r_*$ tends to negative infinity at the event horizon. It is also assumed that the geometry of space time in the general non-stationary black hole is symmetric about $\phi$-axis. According to Refs. [40-46], the generalized tortoise coordinate transformation is defined as
\begin{eqnarray}
r_*&=&r+\frac{1}{2\kappa (u_0,\theta_0,\phi_0)}ln\Big\{\frac{r-r_h(u,\theta,\phi)}{r_h(u,\theta,\phi)}\Big\},\cr
u_*&=&u-u_0,\,\,\,\, \theta_*=\theta-\theta_0,\,\,\,\phi_*=\phi-\phi_0,
\end{eqnarray}
where $u_0$, $\theta_0$, $\phi_0$ are the parameters under the tortoise coordinate transformation. From Eq. (3), we get
\begin{eqnarray}
\frac{\partial}{\partial r}&=&\frac{\partial}{\partial r_*}+\frac{1}{2\kappa(r-r_h)}\frac{\partial}{\partial r_*},\cr
\frac{\partial}{\partial x^{j}}&=&\frac{\partial}{\partial x^{j}_{*}}-\frac{rr_{h,j}}{2\kappa r_h (r-r_h)}\frac{\partial}{\partial r_*},\cr
\frac{\partial^2}{\partial r^2}&=&\frac{[2\kappa(r-r_h)+1]^2}{[2\kappa(r-r_h)]}\frac{\partial^2}{\partial r^2_*}-\frac{1}{2\kappa(r-r_h)^2}\frac{\partial}{\partial r_*},\cr
\frac{\partial^2}{\partial r\partial x^j}&=&\frac{[1+2\kappa(r-r_h)]}{2\kappa(r-r_h)}\frac{\partial^2}{\partial r_*\partial x^j_*}+\frac{r_{h,j}}{2\kappa(r-r_h)^2}\frac{\partial}{\partial r_*}\cr&&-r_hrr_{h,j}\frac{[1+2\kappa(r-r_h)]}{[2r_h\kappa(r-r_h)]^2}\frac{\partial^2}{\partial r^2_*},\cr
\frac{\partial^2}{\partial x^j\partial x^k}&=&\frac{\partial^2}{\partial x^j_*\partial x^k_*}-\frac{rr_{h,j}}{2\kappa(r-r_h)}\frac{\partial^2}{\partial r_*\partial x^k}-\frac{rr_{h,k}}{2\kappa(r-r_h)}\frac{\partial^2}{\partial r_*\partial x^j}
\cr&&+\frac{r^2_hr_{h,j}r_{h,k}}{[2\kappa r_h(r-r_h)]^2}\frac{\partial^2}{\partial r^2_*}
-\frac{r}{2\kappa}\frac{[(r-r_h)r_{h,jk}+r_{h,j}r_{h,k}(2r_h-r)]}{[r_h(r-r_{h}]^2}\frac{\partial}{\partial r_*}.
\end{eqnarray}
Using Eqs. (4) in (2), the horizon equation in general non-stationary black hole is obtained as
\begin{eqnarray}
g^{11}-2g^{1j}r_{h,j}+g^{jk}r_{h,j}r_{h,k}=0,
\end{eqnarray}
where  $r_{h,j}=(\frac{\partial r_{h}}{\partial u}, \frac{\partial r_{h}}{\partial \theta}, \frac{\partial r_{h}}{\partial \phi})$. $r_{h,u}=\frac{\partial r_{h}}{\partial u}$ represents the evaporation rate in general non-stationary black hole near the event horizon. The event horizon is expanded gradually if $\frac{\partial r_{h}}{\partial u}>0$ (absorbing black hole), where, as $\frac{\partial r_{h}}{\partial u}<0$, the event horizon is contracted. In addition, $r_{h,\theta}=\frac{\partial r_{h}}{\partial \theta}$  and $r_{h,\phi}=\frac{\partial r_{h}}{\partial \phi}$  denote the rate of event horizon varying with angles and also describe the rotation effect of non-stationary black hole. $r_{h}$ is the location of event horizon and depends on retarded time $u_0$ and angular coordinates $\theta_0, \phi_0$ and also $\kappa\equiv\kappa(u_0, \theta_0, \phi_0)$ is an adjustable parameter that depends on retarded time and angular coordinates.

3. {\bf Klein-Gordon Equation}.
In this section, the asymptotically behaviour of minimally electromagnetic coupling Klein-Gordon Equation near the black hole will be discussed. The Klein-Gordon equation describes the explicit form of wave equation of the scalar particles with mass $\mu$ in curve space time which is given by
\begin{eqnarray}
\frac{1}{\sqrt{-g}}\Big[\frac{\partial}{\partial x^a}\sqrt{-g}g^{ab}\frac{\partial}{\partial x^b}\Big]\Phi-\mu^2\Phi=0.
\end{eqnarray}
 Using generalized tortoise coordinate transformation to Eq. (6) and subsequently multiplying by the factor $2\kappa r_{h}(r-r_{h})/[r_{h}g^{01}\{1+2\kappa(r-r_{h})\}-rg^{0j}r_{h,j}]$ to both sides of  Eq. (6) and
  finally taking limit near the event horizon as $r\longrightarrow r_h(u_0, \theta_0, \phi_0)$, $u\longrightarrow u_0$, $\theta\longrightarrow \theta_0$ and $\phi\longrightarrow \phi_0$, the second order form of wave equation is obtained as follows
 \begin{eqnarray}
 &&A\frac{\partial^2\Phi}{\partial r^2_*}+2\frac{\partial^2\Phi}{\partial r_*\partial u_*}+A_1\frac{\partial^2\Phi}{\partial r_*\partial\theta_*}+A_2\frac{\partial^2\Phi}{\partial r_*\partial\phi_*}+A_3\frac{\partial\Phi}{\partial r_*}=0,
\end{eqnarray}
where
\begin{eqnarray}
A&=& \frac{g^{11}r^{2}_{h}[1+2\kappa(r-r_{h})]^2-2g^{1j}rr_{h}r_{h,\phi}[1+2\kappa(r-r_{h})]}{2\kappa r_{h}(r-r_{h})[r_{h}g^{01}\{1+2\kappa(r-r_{h}\}-rg^{0j}r_{h,j}]},\cr
 A_1&=&2\frac{g^{12}-g^{2j}r_{h,j}}{(g^{01}-g^{0j}r_{h,j})},\cr
A_2&=&2\frac{g^{13}-g^{3j}r_{h,j}}{(g^{01}-g^{0j}r_{h,j})},\cr
A_3&=&-\frac{\frac{\partial g^{11}}{\partial r}-2\frac{\partial g^{1j}}{\partial r}r_{h,j}+g^{jk}r_{h,jk}}{(g^{01}-g^{0j}r_{h,j})}+\frac{g^{jk}r_{h,j}r_{h,k}}{r^2_{h}(g^{01}-g^{0j}r_{h,j})}(r_h-1)\cr&&
 +\frac{1}{\sqrt{-g}(g^{01}-g^{0j}r_{h,j})}[(\sqrt{-g}_{,\nu}g^{1\nu}+\sqrt{-g}g^{1\nu}_{,\nu})
 -(\sqrt{-g}_{,\nu}g^{0\nu}+\sqrt{-g}g^{0\nu}_{,\nu})\cr&&-(\sqrt{-g}_{,\nu}g^{2\nu}+\sqrt{-g}g^{2\nu}_{,\nu})
 -(\sqrt{-g}_{,\nu}g^{3\nu}+\sqrt{-g}g^{3\nu}_{,\nu})].
\end{eqnarray}
By adjusting parameter $\kappa$, the coefficient of $\frac{\partial^2\Phi}{\partial r^2_{*}}$ is assumed to be unity near the event horizon, we get

\begin{eqnarray}
&&\lim_{r\longrightarrow r_h}\frac{1}{2\kappa r_{h}(r-r_{h})}[g^{11}r^{2}_{h}\{1+2\kappa(r-r_{h})\}^2-2g^{1j}rr_{h}r_{h,\phi}\{1+2\kappa(r-r_{h})\}\cr&&
+r^2g^{jk}r_{h,j}r_{h,k}]=r_{h}(g^{01}-g^{0j}r_{h,j}).
\end{eqnarray}
It is also observed that, in left hand side of Eq. (9), both numerator and denominator tend to zero near the event horizon $r=r_h$. Hence Eq. (9) is an indeterminate form of $0/0$. Using L' Hospital rule and using Eq. (5), the surface gravity is obtained from the Klein-Gordon scalar particles as follows
\begin{eqnarray}
\kappa=\frac{\frac{\partial g^{11}}{\partial r}-2\frac{\partial g^{1j}}{\partial r}r_{h,j}
+\frac{\partial g^{jk}}{\partial r}r_{h,j}r_{h,k}}{2[g^{01}-2g^{11}+(2g^{1j}-g^{0j})r_{h,j}]}
+\frac{g^{jk}r_{h,j}r_{h,k}-g^{1j}r_{h,j}}{r_{h}[g^{01}-2g^{11}+(2g^{1j}-g^{0j})r_{h,j}]}\cdot
\end{eqnarray}

4. {\bf Maxwell's electromagnetic field equations.}

To write the explicit form of Maxwell's electromagnetic field equations in Newman-Penrose formalism [47], the following complex null tetrad vectors $\ell$, $n$, $m$ and $\bar{m}$ are chosen at each point in four dimensional space, where $\ell$ and $n$ are a pair of real null tetrad vectors and $m$ and $\bar{m}$ are a pair of complex null tetrad vectors. They are required to satisfy the following conditions
\begin{eqnarray}
\ell_{\nu}\ell^{\nu}&=&n_{\nu}n^{\nu}=m_{\nu}m^{\nu}=\bar{m}_{\nu}\bar{m}^{\nu}=0,\cr
\ell_{\nu}n^{\nu}&=&-m_{\nu}\bar{m}^{\nu}=1,\cr
\ell_{\nu}m^{\nu}&=&\ell_{\nu}\bar{m}^{\nu}=n_{\nu}m^{\nu}=n_{\nu}\bar{m}^{\nu}=0,\cr
g^{\mu\nu}&=&\ell^{\mu}n^{\nu}+\ell^{\nu}n^{\mu}-m^{\mu}\bar{m}^{\nu}-m^{\nu}\bar{m}^{\mu},\cr
g_{\mu\nu}&=&\ell_{\mu}n_{\nu}+\ell_{\nu}n_{\mu}-m_{\mu}\bar{m}_{\nu}-m_{\nu}\bar{m}_{\mu}
\end{eqnarray}
and corresponding directional derivatives are given by
\begin{eqnarray}
D&=&\ell^{\nu}\frac{\partial}{\partial x^{\mu}},\cr
\Delta&=&n^{\nu}\frac{\partial}{\partial x^{\mu}},\cr
\delta&=&m^{\nu}\frac{\partial}{\partial x^{\mu}},\cr
\bar{\delta}&=&\bar{m}^{\nu}\frac{\partial}{\partial x^{\mu}}.
\end{eqnarray}
The dynamical behavior of spin-1 particles in curve space time is given by four coupled Maxwell's electromagnetic field equations expressed in Newman-Penrose formalism [48] as follows
\begin{eqnarray}
D\phi_1-\bar{\delta}\phi_0-(\pi-2\alpha)\phi_0 -2\rho\phi_1+\kappa\phi_2=0,\cr
\delta\phi_2-\nabla\phi_1+\nu \phi_0 -2\mu\phi_1 -(\tau-2\beta)\phi_2=0, \cr
\delta \phi_1-\nabla \phi_0-(\mu-2\gamma)\phi_0-2\tau \phi_1+\sigma \phi_2=0,\cr
D\phi_2-\bar{\delta}\phi_1+\lambda \phi_0-2\pi\phi_1-(\rho-2\varepsilon)\phi_2=0,
\end{eqnarray}
where $\phi_0$,  $\phi_1$ and $\phi_2$ are the four components of Maxwell's spinor in the Newman-Penrose formalism. $\epsilon, \rho, \pi, \alpha, \beta, \mu, \tau$ and $\gamma$ are spin coefficients introduced by Newman and Penrose, they are given by

\begin{eqnarray}
\rho&=&\ell_{\mu;\nu}m^\mu\bar{m}^\nu,\cr
\pi &=&-n_{\mu;\nu}\bar{m}^\mu \ell^\nu,\cr
\tau &=&\ell_{\mu;\nu}m^\mu n^\nu,\cr
\alpha&=&\frac{1}{2}(\ell_{\mu;\nu}n^\mu\bar{m}^\nu-m_{\mu;\nu}\bar{m}^\mu\bar{m}^\nu),\cr
\beta&=&\frac{1}{2}(\ell_{\mu;\nu}n^\mu m^\nu-m_{\mu;\nu}\bar{m}^\mu m^\nu),\cr
\gamma&=&\frac{1}{2}(\ell_{\mu;\nu}n^\mu n^\nu-m_{\mu;\nu}\bar{m}^\mu n^\nu),\cr
\epsilon&=&\frac{1}{2}(\ell_{\mu;\nu}n^\mu \ell^\nu-m_{\mu;\nu}\bar{m}^\mu \ell^\nu),
\end{eqnarray}
where $\bar{\alpha}$, $\bar{\beta}$, $\bar{\gamma}$, $\bar{\tau}$, $\bar{\epsilon}$, $\bar{\pi}$, $\bar{\mu}$ and $\bar{\rho}$ are complex conjugates of $\alpha$, $\beta$, $\gamma$, $\tau$, $\epsilon$, $\pi$, $\mu$ and $\rho$. From Eqs. (13), the three second order form of Maxwell's equations for ($\phi_0$,  $\phi_1$,  $\phi_2$) components are given by
\begin{eqnarray}
D\nabla\phi_0-\delta\bar{\delta}\phi_0+(\mu-2\gamma)D\phi_0-(\pi-2\alpha)\delta\phi_0-2\rho\nabla\phi_0+2\tau\bar{\delta}\phi_0\\+
(m^\nu\ell^\mu_{,\nu}-\bar{m}^{\mu}_{\nu}\ell^\nu)\frac{\partial \phi_1}{\partial x^{\mu}}+\kappa\nabla\phi_2-\sigma\bar{\delta}\phi_2=0.\nonumber
\end{eqnarray}
\begin{eqnarray}
\nabla D\phi_1-\bar{\delta}\delta\phi_1-2\rho\nabla\phi_1+(\mu-2\gamma)D\phi_1+2\tau\bar{\delta}\phi_1-(\pi-2\alpha)\delta \phi_1\\+(\bar{m}^\nu n^\mu_{,\nu}-n^{\nu}\bar{m}^\mu_{,\nu})\frac{\partial \phi_0}{\partial x^{\mu}}+\kappa\nabla\phi_2-\sigma\bar{\delta}\phi_2=0.\nonumber
\end{eqnarray}
 \begin{eqnarray}
\bar{\delta}\delta\phi_2-\nabla D\phi_2-(\tau-2\beta)\bar{\delta}\phi_2+2\pi\delta\phi_2+(\rho-2\epsilon)\nabla\phi_2-2\mu D\phi_2\\+(\bar{m}^\mu_{\nu} n^{\nu}-n^{\mu}_{,\nu}\bar{m}^\nu)\frac{\partial \phi_1}{\partial x^{\mu}}-\lambda\nabla\phi_0+\nu\bar{\delta}\phi_0=0.\nonumber
\end{eqnarray}
 Refs.  [49-50] have shown that Eqs. (13) cannot be decoupled except only for the stationary black hole space time. For studying the thermal radiation in general non-stationary black hole, the asymptotic behavior of the first order and second order form of Eq. $(13)$ near the event horizon $r=r_h$ will be considered. Then, after taking the limit as $r\longrightarrow r_h(u_0, \theta_0, \phi_0)$, $u\longrightarrow u_0$, $\theta\longrightarrow \theta_0$ and $\phi\longrightarrow \phi_0$, the first order form of Maxwell's equations near the event horizon are as follows:
\begin{eqnarray}
&&(\ell^1-\ell^jr_{h,j})\frac{\partial \phi_1}{\partial r_*}-(\bar{m}^{1}-\bar{m}^{j}r_{h,j})\frac{\partial \phi_0}{\partial r_*}=0,\cr
&&(m^1-m^jr_{h,j})\frac{\partial \phi_2}{\partial r_*}-(n^1-n^{j}r_{h,j})\frac{\partial \phi_1}{\partial r_*}=0,\cr
&&(m^1-m^kr_{h,k})\frac{\partial \phi_1}{\partial r_*}-(n^1-n^{j}r_{h,k})\frac{\partial \phi_0}{\partial r_*}=0,\cr
&&(\ell^1-\ell^kr_{h,k})\frac{\partial \phi_2}{\partial r_*}-(\bar{m}^{1}-\bar{m}^{k}r_{h,k})\frac{\partial \phi_1}{\partial r_*}=0.
\end{eqnarray}
  We assume that the three derivatives $\partial \phi_0/\partial r_{*}$, $\partial \phi_1/\partial r_{*}$ and $\partial \phi_2/\partial r_{*}$ in Eqs. (18) are nonzero. Then non-trivial solutions for $\phi_0$, $\phi_1$ and $\phi_2$ can be obtained if the determinant of their coefficients is zero, which will give the horizon equation like the null surface condition (5). The importance of Eqs. (18) is to eliminate the crossing terms involved in the second order form of Maxwell's equations near the event horizon.

 Utilizing the generalized coordinate transformation (3) to Eqs. (15), (16) and (17) and subsequently multiplying by the factor $2\kappa r_{h}(r-r_{h})/[r_{h}g^{01}\{1+2\kappa(r-r_{h})\}-rg^{0j}r_{h,j}]$ to both sides of three second order equations for the coefficients $\frac{\partial^2 \phi_0}{\partial r_*\partial u_*}$, $\frac{\partial^2 \phi_1}{\partial r_*\partial u_*}$ and $\frac{\partial^2 \phi_2}{\partial r_*\partial u_*}$ to be $2$ and finally taking the limit of $r\longrightarrow r_h(u_0, \theta_0, \phi_0)$, $u\longrightarrow u_0$, $\theta\longrightarrow \theta_0$ and $\phi\longrightarrow \phi_0$, then the three second order form of Maxwell's electromagnetic field equations near the event horizon can be expressed as follows
\begin{eqnarray}
&&I\frac{\partial^2\phi_0}{\partial r^2_*}+2\frac{\partial^2 \phi_0}{\partial r_*\partial u_*}+B_1\frac{\partial^2 \phi_0}{\partial r_*\partial\theta_*}+B_2\frac{\partial^2 \phi_0}{\partial r_*\partial\phi_*}+(B_3+2iB_4)\frac{\partial \phi_0}{\partial r_*}=0,
\end{eqnarray}
where
\begin{eqnarray}
I&=& \frac{g^{11}r^{2}_{h}\{1+2\kappa(r-r_{h})\}^2-2g^{1j}rr_{h}r_{h,\phi}\{1+2\kappa(r-r_{h})\}}{2\kappa r_{h}(r-r_{h})[r_{h}g^{01}\{1+2\kappa(r-r_{h}\}-rg^{0j}r_{h,j}]},\cr
B_1&=&2\frac{(g^{12}-g^{2j}r_{h,j})}{(g^{01}-g^{0j}r_{h,j})},\cr
B_2&=&2\frac{(g^{13}-g^{3j}r_{h,j})}{(g^{01}-g^{0j}r_{h,j})},\cr
B_3&=&-\frac{\frac{\partial g^{11}}{\partial r}-2\frac{\partial g^{1j}}{\partial r}r_{h,j}+g^{jk}r_{h,jk}}{(g^{01}-g^{0j}r_{h,j})}+\frac{g^{jk}r_{h,j}r_{h,k}}{r^2_{h}(g^{01}-g^{0j}r_{h,j})}(r_h-1),\cr
B_4&=&\frac{-i}{(g^{01}-g^{0j}r_{h,j})}[\ell^\nu(n^1_{,\nu}-n^j_{,\nu}r_{h,j})-m^\nu(\ell^1_{,\nu}-\ell^j_{,\nu}r_{h,j})
-2\rho(n^1-n^jr_{h,j})\cr&&+(\mu-2\gamma)(\ell^1_{,\nu}-\ell^j_{,\nu}r_{h,j})-(\pi-2\alpha)(m^1-m^jr_{h,j})
+2\tau(\bar{m}^1-\bar{m}^jr_{h,j})\cr&&
+\frac{(\bar{m}^1-\bar{m}^jr_{h,j})}{(\ell^1-\ell^jr_{h,j})}\{m^\nu(\ell^1_{,\nu}-\ell^j_{,\nu}r_{h,j})
-\ell^\nu(\bar{m}^1_{,\nu}-\bar{m}^jr_{h,j})\}\cr&&
-\{\sigma(\bar{m}^1-\bar{m}^jr_{h,j})
-\kappa(n^1-n^jr_{h,j})\}\frac{(n^1-n^jr_{h,j})(\bar{m}^1-\bar{m}^jr_{h,j})}{(m^1-m^jr_{h,j})(\ell^1-\ell^jr_{h,j})}]
\end{eqnarray}
and
\begin{eqnarray}
&&I\frac{\partial^2\phi_1}{\partial r^2_*}+2\frac{\partial^2 \phi_1}{\partial r_*\partial u_*}+B_1\frac{\partial^2 \phi_1}{\partial r_*\partial\theta_*}+B_2\frac{\partial^2 \phi_1}{\partial r_*\partial\phi_*}+(B_3+2i\tilde{B}_4)\frac{\partial \phi_1}{\partial r_*}=0,
\end{eqnarray}
where
\begin{eqnarray}
\tilde{B}_4&=&\frac{-i}{(g^{01}-g^{0j}r_{h,j})}[-2\rho(n^1-n^jr_{h,j})+(\mu-2\gamma)(\ell^1-\ell^jr_{h,j})
+2\tau(\bar{m}^1-\bar{m}^jr_{h,j})\cr&&-(\pi-2\alpha)(m^1-m^jr_{h,j})
+n^\nu(\ell^1_{,\nu}-\ell^j_{,\nu}r_{h,j})-\bar{m}^\nu(m^1_{,\nu}-m^j_{,\nu}r_{h,j})\cr&&
+\frac{(\ell^1-\ell^jr_{h,j})}{(\bar{m}^1-\bar{m}^jr_{h,j})}\{\bar{m}^\nu(n^1_{,\nu}-n^j_{,\nu}r_{h,j})
-n^\nu(\bar{m}^1_{,\nu}-\bar{m}^j_{,\nu}r_{h,j})\}\cr&&
+\{-\sigma(\bar{m}^1-\bar{m}^jr_{h,j})
+\kappa(n^1-n^jr_{h,j})\}\frac{(n^1-n^jr_{h,j})}{(m^1-m^jr_{h,j})}],
\end{eqnarray}
and
\begin{eqnarray}
&&I\frac{\partial^2\phi_2}{\partial r^2_*}+2\frac{\partial^2 \phi_2}{\partial r_*\partial u_*}+B_1\frac{\partial^2 \phi_2}{\partial r_*\partial\theta_*}+B_2\frac{\partial^2 \phi_2}{\partial r_*\partial\phi_*}+(B_3+2i\bar{B}_4)\frac{\partial \phi_2}{\partial r_*}=0,
\end{eqnarray}
where
\begin{eqnarray}
\bar{B}_4&=&\frac{-i}{(g^{01}-g^{0j}r_{h,j})}[n^\nu(\ell^1_{,\nu}-\ell^j_{,\nu}r_{h,j})-\bar{m}^\nu(m^1_{,\nu}-m^j_{,\nu}r_{h,j})
\cr&&-2\pi(m^1_{,\nu}-m^j_{,\nu}r_{h,j})+(\tau-2\beta)(\bar{m}^1_{,\nu}-\bar{m}^j_{,\nu})-2(\rho-2\epsilon)(n^1_{,\nu}-n^j_{,\nu}r_{h,j})
\cr&&+2\mu(\ell^1_{,\nu}-\ell^j_{,\nu}r_{h,j})
+\frac{(\ell^1-\ell^jr_{h,j})} {(\bar{m}^1-\bar{m}^jr_{h,j})}\frac{(m^1-m^jr_{h,j})}{(n^1-n^jr_{h,j})}\{\lambda(n^1_{,\nu}-n^j_{,\nu}r_{h,j})\cr&&
-\nu(\bar{m}^1_{,\nu}-\bar{m}^j_{,\nu}r_{h,j})\}-2\{n^\nu(\bar{m}^1-\bar{m}^jr_{h,j})
\cr&&-\bar{m}^\nu(n^1_{,\nu}-n^j_{,\nu}r_{h,j})\}\frac{(m^1-m^jr_{h,j})}{(n^1-n^jr_{h,j})}].
\end{eqnarray}
We assume the value of $I$ approaches to unity near the event horizon, then we get
\begin{eqnarray}
\lim_{r\longrightarrow r_h}\frac{g^{11}r^{2}_{h}\{1+2\kappa(r-r_{h})\}^2-2g^{1j}rr_{h}r_{h,\phi}\{1+2\kappa(r-r_{h})\}}{2\kappa r_{h}(r-r_{h})[r_{h}g^{01}\{1+2\kappa(r-r_{h})\}-rg^{0j}r_{h,j}]}=1,
\end{eqnarray}
which is an indeterminate form of $0/0$ and applying L' Hospital rule, the surface gravity due to the Dirac particles is given by
\begin{eqnarray}
\kappa=\frac{\frac{\partial g^{11}}{\partial r}-2\frac{\partial g^{1j}}{\partial r}r_{h,j}
+\frac{\partial g^{jk}}{\partial r}r_{h,j}r_{h,k}}{2[g^{01}-2g^{11}+(2g^{1j}-g^{0j})r_{h,j}]}
+\frac{g^{jk}r_{h,j}r_{h,k}-g^{1j}r_{h,j}}{r_{h}[g^{01}-2g^{11}+(2g^{1j}-g^{0j})r_{h,j}]}\cdot
\end{eqnarray}
which is equal to the surface gravity derived from Klein-Gordon scalar particle given by Eq. (10).

5. {\bf Dirac equations.}
The four couples Dirac equations [51] expressed in Newman-Penrose formalism are
given by
\begin{eqnarray}
&&(D+\epsilon-\rho)f_1 +
(\bar{\delta}+\pi-\alpha)f_2-\frac{i\mu_0g_1}{\sqrt{2}}=0,\cr
&&(\nabla+\mu-\gamma)f_2+(\delta+\beta-\tau)F_1-\frac{i\mu_0g_2}{\sqrt{2}}=0, \cr
&&(\nabla+\bar{\mu}-\bar{\gamma})g_1-(\bar{\delta}+\bar{\beta}-\bar{\tau})g_2-\frac{i\mu_0f_1}{\sqrt{2}}=0,\cr
&&(D+\bar{\epsilon}-\bar{\rho})g_2-(\delta+\bar{\pi}-\bar{\alpha})g_1-\frac{i\mu_0f_2}{\sqrt{2}}=0,
\end{eqnarray}
where $D$, $\nabla$, $\delta$ and $\bar{\delta}$ are the directional derivatives given by Eqs. (12) and  $\epsilon, \rho, \pi, \alpha, \beta, \mu, \tau$ and $\gamma$ are spin coefficients and also $\mu_0$ is the mass of the Dirac particles. $f_1, f_2, g_1$ and $g_2$ are the four components of Dirac spinor in the Newman-Penrose formalism. Eqs. (27) can be decoupled only for the stationary black hole space time. From Eqs. (27) the second order form of Dirac equations for the components $(f_1, f_2)$ are given by
\begin{eqnarray}
-2(\nabla+\bar{\mu}-\bar{\gamma})\times[(D+\epsilon-\rho)f_1+(\bar{\delta}+\phi-\alpha)f_2]
+2(\bar{\delta}+\bar{\beta}-\bar{\tau})\times[(\nabla+\mu-\gamma)f_2\\+(\delta+\beta-\tau)f_1]-\mu^2_0f_1=0.\nonumber
\end{eqnarray}
\begin{eqnarray}
-2(D+\bar{\epsilon}-\bar{\rho})\times[(\nabla+\mu-\gamma)f_2+(\delta+\beta-\tau)f_1]
+2(\delta+\bar{\pi}-\bar{\alpha})\times[(D+\epsilon-\rho)f_1\\+(\bar{\delta}+\pi-\alpha)f_2]-\mu^2_0f_2=0.\nonumber
\end{eqnarray}
Applying generalized tortoise coordinate transformation to Eqs. (27), after taking limit  $r\longrightarrow r_h(u_0, \theta_0, \phi_0)$, $u\longrightarrow u_0$, $\theta\longrightarrow \theta_0$ and $\phi\longrightarrow \phi_0$, the first order form of Dirac equations near the event horizon are given by
\begin{eqnarray}
\frac{\partial f_1}{\partial r_*}&=&\frac{\bar{m}^1-\bar{m}^jr_{h,j}}{\ell^jr_{h,j}-\ell^1}\frac{\partial f_2}{\partial r_*},\cr
\frac{\partial f_2}{\partial r_*}&=&\frac{m^1-m^jr_{h,j}}{n^jr_{h,j}-n^1}\frac{\partial f_1}{\partial r_*},\cr
\frac{\partial g_2}{\partial r_*}&=&\frac{m^1-m^kr_{h,k}}{\ell^1-\ell^k r_{h,k}}\frac{\partial g_1}{\partial r_*},\cr
\frac{\partial g_1}{\partial r_*}&=&\frac{\bar{m}^1-\bar{m}^kr_{h,k}}{n^1-n^kr_{h,k}}\frac{\partial g_2}{\partial r_*}.
\end{eqnarray}
In order to study Hawking thermal radiation from Dirac particles, we should consider the asymptotic behaviour of Eqs. (27) near the black hole horizon.  The non-trivial solutions for $f_1$, $f_2$, $g_1$ and $g_2$ can be obtained if the four derivatives $\partial f_1/\partial r_{*}$, $\partial f_2/\partial r_{*}$, $\partial g_1/\partial r_{*}$ and $\partial g_2/\partial r_{*}$ in Eqs. (30) are nonzero. Eqs. (30) may be used to eliminate the crossing terms appeared in the second order form of Dirac equation near the horizon.

Applying generalized tortoise coordinate transformation to the Eqs. (28) and (29), via some arrangement, multiplying by the factor $2\kappa r_{h}(r-r_{h})/[r_{h}g^{01}\{1+2\kappa(r-r_{h}\}-rg^{0j}r_{h,j}]$ to both sides of the two second order form of Dirac equations for the coefficient $\frac{\partial^2 f_1}{\partial r_*\partial u_*}$ and $\frac{\partial^2 f_2}{\partial r_*\partial u_*}$ to be 2 and taking the limit of $r\longrightarrow r_h(u_0, \theta_0, \phi_0)$, $u\longrightarrow u_0$, $\theta\longrightarrow \theta_0$ and $\phi\longrightarrow \phi_0$, the two second order form of Dirac equations can be written as follows
\begin{eqnarray}
&&Q\frac{\partial^2f_1}{\partial r^2_*}+2\frac{\partial^2 f_1}{\partial r_*\partial u_*}+C_1\frac{\partial^2 f_1}{\partial r_*\partial\theta_*}+C_2\frac{\partial^2 f_1}{\partial r_*\partial\phi_*}+(C_3+2iC_4)\frac{\partial f_1}{\partial r_*}=0,
\end{eqnarray}
where
\begin{eqnarray}
Q&=& \frac{g^{11}r^{2}_{h}\{1+2\kappa(r-r_{h})\}^2-2g^{1j}rr_{h}r_{h,\phi}\{1+2\kappa(r-r_{h})\}}{2\kappa r_{h}(r-r_{h})[r_{h}g^{01}\{1+2\kappa(r-r_{h})\}-rg^{0j}r_{h,j}]},\cr
C_1&=&2\frac{(g^{12}-g^{2j}r_{h,j})}{(g^{01}-g^{0j}r_{h,j})},\cr
C_2&=&2\frac{(g^{13}-g^{3j}r_{h,j})}{(g^{01}-g^{0j}r_{h,j})},\cr
C_3&=&-\frac{\frac{\partial g^{11}}{\partial r}-2\frac{\partial g^{1j}}{\partial r}r_{h,j}+g^{jk}r_{h,jk}}{(g^{01}-g^{0j}r_{h,j})}+\frac{g^{jk}r_{h,j}r_{h,k}}{r^2_{h}(g^{01}-g^{0j}r_{h,j})}(r_h-1),\cr
C_4&=&\frac{-i}{(g^{01}-g^{0j}r_{h,j})}[-(\beta-\tau+\bar{\beta}-\bar{\tau})(\bar{m}^1-\bar{m}^jr_{h,j})
+(\epsilon-\rho)(n^1-n^jr_{h,j})\cr&&+(\bar{\mu}-\bar{\gamma})(\ell^1-\ell^jr_{h,j})
-\frac{(\ell^1-\ell^jr_{h,j})}{(\bar{m}^1-\bar{m}^jr_{h,j})}\{-\bar{m}^\nu(n^1_{,\nu}-n^j_{,\nu})\cr&&
+n^\nu(\bar{m}^1-\bar{m}^j_{,\nu})\}-(\bar{\mu}-\bar{\gamma}-\mu+\gamma)(\ell^1-\ell^jr_{h,j})
-(\pi-\alpha-\bar{\beta}+\bar{\tau})\cr&&\times\frac{(\ell^1-\ell^jr_{h,j})^2}{(\bar{m}^1-\bar{m}^jr_{h,j})}]
\end{eqnarray}
and
\begin{eqnarray}
&&Q\frac{\partial^2f_2}{\partial r^2_*}+2\frac{\partial^2 f_2}{\partial r_*\partial u_*}+C_1\frac{\partial^2 f_2}{\partial r_*\partial\theta_*}+C_2\frac{\partial^2 f_2}{\partial r_*\partial\phi_*}+(C_3+2i\bar{C}_4)\frac{\partial f_2}{\partial r_*}=0,
\end{eqnarray}
where
\begin{eqnarray}
\bar{C}_4&=&\frac{-i}{g^{01}-g^{0j}r_{h,j}}[(\mu-\gamma)(\ell^1-\ell^jr_{h,j})-(\pi-\alpha)(m^1-m^jr_{h,j})
\cr&&+\ell^\nu(n^1_{,\nu}-n^j_{,\nu})
+(\bar{\epsilon}-\bar{\rho})(n^1-n^jr_{h,j})+\ell^\nu(n^1_{,\nu}-n^j_{,\nu})-m^\nu(\bar{m}^1-\bar{m}^j_{,\nu})
\cr&&-(\bar{\epsilon}-\bar{\rho}-\epsilon+\rho)(n^1_{,\nu}-n^j_{,\nu})
-\ell^\nu\frac{(m^1_{,\nu}-m^j_{,\nu})}{(\bar{m}^1-\bar{m}^j_{h,j})}(n^1-n^jr_{h,j})\cr&&+\{m^\nu
\frac{(\ell^1_{,\nu}-\ell^j_{,\nu})}{(m^1-m^jr_{h,j})}-(\beta-\tau-\bar{\pi}
+\bar{\alpha})\frac{(\ell^1-\ell^jr_{h,j})}{(m^1-m^jr_{h,j})}\}(n^1-n^jr_{h,j})],
\end{eqnarray}
when $Q$ approaches to unity, we obtain
\begin{eqnarray}
\lim_{r\longrightarrow r_h} \frac{g^{11}r^{2}_{h}\{1+2\kappa(r-r_{h})\}^2-2g^{1j}rr_{h}r_{h,\phi}\{1+2\kappa(r-r_{h})\}}{2\kappa r_{h}(r-r_{h})[r_{h}g^{01}\{1+2\kappa(r-r_{h})\}-rg^{0j}r_{h,j}]}=1.
\end{eqnarray}
The above Eq. (35) is a $0/0$ indeterminate form. By applying L' Hospital rule near the black hole event horizon, the surface gravity due to the Dirac particle, is given by
\begin{eqnarray}
\kappa=\frac{\frac{\partial g^{11}}{\partial r}-2\frac{\partial g^{1j}}{\partial r}r_{h,j}
+\frac{\partial g^{jk}}{\partial r}r_{h,j}r_{h,k}}{2[g^{01}-2g^{11}+(2g^{1j}-g^{0j})r_{h,j}]}
+\frac{g^{jk}r_{h,j}r_{h,k}-g^{1j}r_{h,j}}{r_{h}[g^{01}-2g^{11}+(2g^{1j}-g^{0j})r_{h,j}]}\cdot
\end{eqnarray}
which is the same as Eqs. (10) and (26), the surface gravities derived from Klein-Gordon Scalar particles, and Maxwell's electromagnetic field equations.

6. {\bf Thermal radiation spectrum.} To investigate the thermal radiation spectrum in general non-stationary black hole, we combine the second order form of Klein-Gordon equation (7), the three second order form Maxwell's equations (19), (21), (23) and the two second order form of Dirac equations (31), (33) as single form of wave equation near the event horizon $r=r_h$ as follows
\begin{eqnarray}
&&\frac{\partial^2\Psi}{\partial r^2_*}+2\frac{\partial^2\Psi}{\partial r_*\partial u_*}+L_1\frac{\partial^2\Psi}{\partial r_*\partial\theta_*}+L_2\frac{\partial^2\Psi}{\partial r_*\partial\phi_*}+(L_3+2iL_4)\frac{\partial\Psi}{\partial r_*}=0,
\end{eqnarray}
where
\begin{eqnarray}
L_1&=&2\frac{(g^{12}-g^{2j}r_{h,j})}{(g^{01}-g^{0j}r_{h,j})},\cr
L_2&=&2\frac{(g^{13}-g^{3j}r_{h,j})}{(g^{01}-g^{0j}r_{h,j})}.
\end{eqnarray}
The Eq. (37) may be assumed as standard form of wave equation in general non-stationary black hole near the horizon $r=r_h$. It includes Klein-Gordon equation, Maxwell's electromagnetic field equations and Dirac equations with different co-efficient of constant terms.

For example, when $(\Psi=\Phi)$ for the Klein-Gordon equation, the Eq. (37) gives the following constants
\begin{eqnarray}
L_3&=&-\frac{\frac{\partial g^{11}}{\partial r}-2\frac{\partial g^{1j}}{\partial r}r_{h,j}+g^{jk}r_{h,jk}}{(g^{01}-g^{0j}r_{h,j})}+\frac{g^{jk}r_{h,j}r_{h,k}}{r^2_{h}(g^{01}-g^{0j}r_{h,j})}(r_h-1)\cr&&
 +\frac{1}{\sqrt{-g}(g^{01}-g^{0j}r_{h,j})}[(\sqrt{-g}_{,\nu}g^{1\nu}+\sqrt{-g}g^{1\nu}_{,\nu})
 -(\sqrt{-g}_{,\nu}g^{0\nu}+\sqrt{-g}g^{0\nu}_{,\nu})\cr&&-(\sqrt{-g}_{,\nu}g^{2\nu}+\sqrt{-g}g^{2\nu}_{,\nu})
 -(\sqrt{-g}_{,\nu}g^{3\nu}+\sqrt{-g}g^{3\nu}_{,\nu})],\cr
L&=&0.
\end{eqnarray}

 For Maxwell's electromagnetic equations $(\Psi=\phi_0)$, the constant terms are
\begin{eqnarray}
L_3&=&-\frac{\frac{\partial g^{11}}{\partial r}-2\frac{\partial g^{1j}}{\partial r}r_{h,j}+g^{jk}r_{h,jk}}{(g^{01}-g^{0j}r_{h,j})}+\frac{g^{jk}r_{h,j}r_{h,k}}{r^2_{h}(g^{01}-g^{0j}r_{h,j})}(r_h-1),\cr
L_4&=&\frac{-i}{(g^{01}-g^{0j}r_{h,j})}[\ell^\nu(n^1_{,\nu}-n^j_{,\nu}r_{h,j})-m^\nu(\ell^1_{,\nu}-\ell^j_{,\nu}r_{h,j})
-2\rho(n^1-n^jr_{h,j})\cr&&+(\mu-2\gamma)(\ell^1_{,\nu}-\ell^j_{,\nu}r_{h,j})-(\pi-2\alpha)(m^1-m^jr_{h,j})
+2\tau(\bar{m}^1-\bar{m}^jr_{h,j})\cr&&
+\frac{(\bar{m}^1-\bar{m}^jr_{h,j})}{(\ell^1-\ell^jr_{h,j})}\{m^\nu(\ell^1_{,\nu}-\ell^j_{,\nu}r_{h,j})
-\ell^\nu(\bar{m}^1_{,\nu}-\bar{m}^jr_{h,j})\cr&&
-\{\sigma(\bar{m}^1-\bar{m}^jr_{h,j})
-\kappa(n^1-n^jr_{h,j})\}\frac{(n^1-n^jr_{h,j})(\bar{m}^1-\bar{m}^jr_{h,j})}{(m^1-m^jr_{h,j})(\ell^1-\ell^jr_{h,j})}].
\end{eqnarray}
For $(\Psi=\phi_1)$
\begin{eqnarray}
L_3&=&-\frac{\frac{\partial g^{11}}{\partial r}-2\frac{\partial g^{1j}}{\partial r}r_{h,j}+g^{jk}r_{h,jk}}{(g^{01}-g^{0j}r_{h,j})}+\frac{g^{jk}r_{h,j}r_{h,k}}{r^2_{h}(g^{01}-g^{0j}r_{h,j})}(r_h-1),\cr
L_4&=&\frac{-i}{(g^{01}-g^{0j}r_{h,j})}[-2\rho(n^1-n^jr_{h,j})+(\mu-2\gamma)(\ell^1-\ell^jr_{h,j})
+2\tau(\bar{m}^1-\bar{m}^jr_{h,j})\cr&&-(\pi-2\alpha)(m^1-m^jr_{h,j})
+n^\nu(\ell^1_{,\nu}-\ell^j_{,\nu}r_{h,j})-\bar{m}^\nu(m^1_{,\nu}-m^j_{,\nu}r_{h,j})\cr&&
+\frac{(\ell^1-\ell^jr_{h,j})}{(\bar{m}^1-\bar{m}^jr_{h,j})}\{\bar{m}^\nu(n^1_{,\nu}-n^j_{,\nu}r_{h,j})
-n^\nu(\bar{m}^1_{,\nu}-\bar{m}^j_{,\nu}r_{h,j})\}\cr&&
+\{-\sigma(\bar{m}^1-\bar{m}^jr_{h,j})
+\kappa(n^1-n^jr_{h,j})\}\frac{(n^1-n^jr_{h,j})}{(m^1-m^jr_{h,j})}].
\end{eqnarray}
Lastly  for $(\Psi=\phi_2)$, we get
\begin{eqnarray}
L_3&=&-\frac{\frac{\partial g^{11}}{\partial r}-2\frac{\partial g^{1j}}{\partial r}r_{h,j}+g^{jk}r_{h,jk}}{(g^{01}-g^{0j}r_{h,j})}+\frac{g^{jk}r_{h,j}r_{h,k}}{r^2_{h}(g^{01}-g^{0j}r_{h,j})}(r_h-1),\cr
L_4&=&\frac{-i}{(g^{01}-g^{0j}r_{h,j})}[n^\nu(\ell^1_{,\nu}-\ell^j_{,\nu}r_{h,j})-\bar{m}^\nu(m^1_{,\nu}-m^j_{,\nu}r_{h,j})
\cr&&-2\pi(m^1_{,\nu}-m^j_{,\nu}r_{h,j})+(\tau-2\beta)(\bar{m}^1_{,\nu}-\bar{m}^jr_{h,j})-2(\rho-2\epsilon)(n^1_{,\nu}-n^j_{,\nu}r_{h,j})
\cr&&+2\mu(\ell^1_{,\nu}-\ell^j_{,\nu}r_{h,j})
+\frac{(\ell^1-\ell^jr_{h,j})} {(\bar{m}^1-\bar{m}^jr_{h,j})}\frac{(m^1-m^jr_{h,j})}{(n^1-n^jr_{h,j})}\{\lambda(n^1_{,\nu}-n^j_{,\nu}r_{h,j})\cr&&
-\nu(\bar{m}^1_{,\nu}-\bar{m}^j_{,\nu}r_{h,j})\}-2\{n^\nu(\bar{m}^1-\bar{m}^jr_{h,j})
\cr&&-\bar{m}^\nu(n^1_{,\nu}-n^j_{,\nu}r_{h,j})\}\frac{(m^1-m^jr_{h,j})}{(n^1-n^jr_{h,j})}].
\end{eqnarray}
Similarly for Dirac particles when $(\Psi=f_1)$, Eq. (37) gives the following constant terms
\begin{eqnarray}
L_3&=&-\frac{\frac{\partial g^{11}}{\partial r}-2\frac{\partial g^{1j}}{\partial r}r_{h,j}+g^{jk}r_{h,jk}}{(g^{01}-g^{0j}r_{h,j})}+\frac{g^{jk}r_{h,j}r_{h,k}}{r^2_{h}(g^{01}-g^{0j}r_{h,j})}(r_h-1),\cr
L_4&=&\frac{-i}{(g^{01}-g^{0j}r_{h,j})}[-(\beta-\tau+\bar{\beta}-\bar{\tau})(\bar{m}^1-\bar{m}^jr_{h,j})
+(\epsilon-\rho)(n^1-n^jr_{h,j})\cr&&+(\bar{\mu}-\bar{\gamma})(\ell^1-\ell^jr_{h,j})
-\frac{(\ell^1-\ell^jr_{h,j})}{(\bar{m}^1-\bar{m}^jr_{h,j})}\{-\bar{m}^\nu(n^1_{,\nu}-n^j_{,\nu})\cr&&
+n^\nu(\bar{m}^1-\bar{m}^j_{,\nu})\}-(\bar{\mu}-\bar{\gamma}-\mu+\gamma)(\ell^1-\ell^jr_{h,j})
-(\pi-\alpha-\bar{\beta}+\bar{\tau})\cr&&\times\frac{(\ell^1-\ell^jr_{h,j})^2}{(\bar{m}^1-\bar{m}^jr_{h,j})}],
\end{eqnarray}
for $(\Psi=f_2)$, we obtain
\begin{eqnarray}
L_3&=&-[\frac{\frac{\partial g^{11}}{\partial r}-2\frac{\partial g^{1j}}{\partial r}r_{h,j}+g^{jk}r_{h,jk}}{(g^{01}-g^{0j}r_{h,j})}]+\frac{g^{jk}r_{h,j}r_{h,k}}{r^2_{h}(g^{01}-g^{0j}r_{h,j})}(r_h-1),\cr
L_4&=&\frac{-i}{(g^{01}-g^{0j}r_{h,j})}[(\mu-\gamma)(\ell^1-\ell^jr_{h,j})-(\pi-\alpha)(m^1-m^jr_{h,j})
\cr&&+\ell^\nu(n^1_{,\nu}-n^j_{,\nu})
+(\bar{\epsilon}-\bar{\rho})(n^1-n^jr_{h,j})+\ell^\nu(n^1_{,\nu}-n^j_{,\nu})-m^\nu(\bar{m}^1-\bar{m}^j_{,\nu})
\cr&&-(\bar{\epsilon}-\bar{\rho}-\epsilon+\rho)(n^1_{,\nu}-n^j_{,\nu})
-\ell^\nu\frac{(m^1_{,\nu}-m^j_{,\nu})}{(\bar{m}^1-\bar{m}^j_{h,j})}(n^1-n^jr_{h,j})\cr&&+\{m^\nu
\frac{(\ell^1_{,\nu}-\ell^j_{,\nu})}{(m^1-m^jr_{h,j})}-(\beta-\tau-\bar{\pi}
+\bar{\alpha})\frac{(\ell^1-\ell^jr_{h,j})}{(m^1-m^jr_{h,j})}\}(n^1-n^jr_{h,j})].
\end{eqnarray}
 Eq. (37) may be assumed as  second order partial differential equation near the event horizon in general non-stationary black hole since all the coefficients $L_1$, $L_2$, $L_3$ and $L_4$ are constant terms when $r\longrightarrow r_h(u_0, \theta_0, \phi_0)$, $u\longrightarrow u_0$, $\theta\longrightarrow \theta_0$ and $\phi\longrightarrow \phi_0$.

 Using Refs. [29, 30, 36, 52], the variables in Eq. (37) may be separated for the analysis of the field equations as
\begin{eqnarray}
\Psi=R(r_*,u_*)\Theta(u_*,\theta_*,\phi_*)e^{i\omega u_*+iK_\theta+iK_\phi},
\end{eqnarray}
where $\Theta(u_*,\theta_*,\phi_*)$ is an arbitrary real function and $\omega$ is the energy of the particles which depend on tortoise coordinate transformation; $K_{\theta}$, $K_{\phi}$ are components of generalized momentum of scalar particles. And we use $K_\theta=\frac{\partial S}{\partial \theta_*}$, $K_\phi=\frac{\partial S}{\partial \phi_*}$, where $S$ is Hamiltonian function of scalar particles. Using Eq. (45) into Eq. (37) and after separating the variables, the radial and angular parts are given by

\begin{eqnarray}
&&\frac{\partial^2 P}{\partial r^2_*}-(2i\omega-L_1iK_\theta-L_2iK_{\phi}-L_3-i2L_4-\alpha)\frac{\partial P}{\partial r_*}=0,\cr
&&\frac{2\partial T}{\partial u_*}-(\zeta(u_*)-\alpha)T=0,
\end{eqnarray}
where $\alpha$ and $\zeta(u_*)$ are a constant and function of retarded time $u_*$  in variable of separation respectively, where $\zeta(u_*)=2\frac{\frac{\partial\Theta}{\partial u_*}}{\Theta}+C_1\frac{\frac{\partial\Theta}{\partial \theta_*}}{\Theta}-C_2\frac{\frac{\partial\Theta}{\partial \phi_*}}{\Theta}$ and $R(r_*, u_*)=P(r_*)T( u_*)$.

After separation of variables, the radial components of two independent solutions are defined by
\begin{eqnarray}
\Psi^{in}_\omega &\sim& e^{i\omega+iK_\theta+iK_\phi},\cr
\Psi^{out}_\omega&\sim&e^{i\omega+iK_\theta+iK_\phi} e^{2i(\omega-\frac{K_\theta L_1}{2}-\frac{K_\phi L_2}{2}-L_4)r_*}e^{-(\alpha+L_3)r_*},\,\,\,\,\,\,\,\,\,\,\,\,\,\,\,\,\,r>r_h,
\end{eqnarray}
where $\Psi^{in}_\omega$ represents an incoming wave which is analytic at $r=r_h$; $\Psi^{out}_\omega$ denotes an outgoing wave having singularity at the event horizon. Refs. [27-28] indicates that $\Psi^{out}_\omega$ can continue analytically from outside of the event horizon $r=r_h$ into inside by rotating $-\pi$ through lower half of the complex plane as
\begin{eqnarray}
\tilde{\Psi}^{out}_{\omega}&\sim &e^{i\omega+iK_\theta+iK_\phi} e^{\frac{\pi}{\kappa}[\omega-\frac{K_\theta L_1}{2}-\frac{K_\phi L_2}{2}-L_4]}e^{\frac{i\pi}{2\kappa}(\alpha+L_3)}\cr&&\times \Big(\frac{r_h-r}{r_h}\Big)^{\frac{i}{\kappa}(\omega-\frac{K_\theta L_1}{2}-\frac{K_\phi L_2}{2}-L_4)}\Big(\frac{r_h-r}{r_h}\Big)^{\frac{-1}{2\kappa}(\alpha+L_3)},\,\,\,\,\,\,\,r<r_h.
\end{eqnarray}
From Eqs. (47) and (48), one can obtain the relative scattering probability near the event horizon $r=r_h$
\begin{eqnarray}
\Big|\frac{\Psi^{out}_\omega(r>r_h)}{\tilde{\Psi}^{out}_{\omega}(r<r_h)}\Big|^2=e^{-\frac{2\pi}{\kappa}(\omega-\omega_0)},
\end{eqnarray}
where
\begin{eqnarray}
\omega_0=K_\theta\frac{L_1}{2}+K_\phi\frac{L_2}{2}+L_4.
\end{eqnarray}
Following Damour and Ruffini [27] and extended by Sannan [28], the thermal radiation spectrum of Maxwell's electromagnetic field equations (Dirac particles or scalar particles) from general non-stationary black holes is given by
\begin{eqnarray}
N_\omega=\frac{1}{e^{\frac{\omega-\omega_0}{\kappa_B T}}\pm1},
\end{eqnarray}
where $\kappa_B$ is Boltzmann constant and upper positive symbol stands for the Fermi-Dirac distribution and the lower negative symbol corresponds to the Bose-Einstein statistics. Eq. (51) shows that the black hole radiates like a black body. The Hawking temperature is given by
\begin{eqnarray}
T(u_0, \theta_0, \phi_0)&=&\frac{1}{2\pi}\Big[\frac{\frac{\partial g^{11}}{\partial r}-2\frac{\partial g^{1j}}{\partial r}r_{h,j}
+\frac{\partial g^{jk}}{\partial r}r_{h,j}r_{h,k}}{2\{g^{01}-2g^{11}+(2g^{1j}-g^{0j})r_{h,j}\}}
\cr&&+\frac{g^{jk}r_{h,j}r_{h,k}-g^{1j}r_{h,j}}{r_{h}\{g^{01}-2g^{11}+(2g^{1j}-g^{0j})r_{h,j}\}}\Big],
\end{eqnarray}
and the chemical potential by
\begin{eqnarray}
\omega_0&=&K_{\theta}\frac{g^{12}-g^{2j}r_{h,j}}{g^{01}-g^{0j}r_{h,j}}
+K_{\phi}\frac{g^{13}-g^{3j}r_{h,j}}{g^{01}-g^{0j}r_{h,j}}+L_4.
\end{eqnarray}
Integrating the thermal radiation spectra (51) or distribution function over all $\omega$'s the combined form of Hawking flux for Klein-Gordon scalar particles, Maxwell's electromagnetic field equations and Dirac equations can be obtained as follows
\begin{eqnarray}
{\rm Flux}=\frac{1}{\pi}\int^{\infty}_{0}\frac{\omega d\omega}{e^{\frac{2\pi}{\kappa}(\omega-\omega_0)}\pm1}.
\end{eqnarray}
This is an exact result for the energy flux in general non-stationary black hole. If $\omega_0=0$ in Eq. (54), the Hawking flux for fermions is given by
\begin{eqnarray}
{(\rm Flux)}_{\mid{\rm fermions}}=\frac{1}{\pi}\int^{\infty}_{0}\frac{\omega d\omega}{e^{\frac{2\pi\omega}{\kappa}}+1}=\frac{\kappa^2}{48\pi},
\end{eqnarray}
and the Hawking flux for boson is defined by
\begin{eqnarray}
{(\rm Flux)}_{\mid{\rm boson}}=\frac{1}{\pi}\int^{\infty}_{0}\frac{\omega d\omega}{e^{\frac{2\pi\omega}{\kappa}}-1}=\frac{\kappa^2}{24\pi}.
\end{eqnarray}
This results are consistent with already obtained in the literature [53, 54].
From Eq. (52), we observe that $T$ is a function of retarded time and different angles. Hence, it is a distribution of temperature of the thermal radiation near the event horizon $r=r_h$ due to the Klein-Gordon scalar field, the Maxwell's electromagnetic field equations and the Dirac equations in general non-stationary black hole. It has been shown that the constant coefficient $L_4$ appears in the expression of chemical potential and may represent a particular energy term for Maxwell's electromagnetic field and Dirac particles which is absent in the thermal radiation spectrum of other scalar particles.

7. {\bf Non-thermal radiation.}

The relativistic Hamilton-Jacobi equation for the classical action of a particle of mass $\mu_0$ in a curve space time is given by [55]
\begin{eqnarray}
g^{ab}\Big(\frac{\partial\Phi}{\partial x^a}\Big)\Big(\frac{\partial\Phi}{\partial x^b}\Big)-\mu^2_0=0,
\end{eqnarray}
where $\Phi=\Phi(u,r,\theta,\phi)$ is the Hamiltonian principal function. Using Eq. (3) into Eq. (57), we obtain as follows
\begin{eqnarray}
\frac{G}{2\kappa(r-r_h)r_h}(\frac{\partial S}{\partial r_*})^2-2D(\frac{\partial S}{\partial r_*})+2r_h\kappa(r-r_h)Y=0,
\end{eqnarray}
where
\begin{eqnarray}
G&=&g^{11}r^2_{h}[1+2\kappa(r-r_h)]^2+r^2g^{jk}r_{h,j}r_{h,k}-2rr_{h}[1+2\kappa(r-r_h)]r_{h,j}g^{1j},\cr
D&=&rg^{jk}(\frac{\partial S}{\partial x^j_*})-r_h(\frac{\partial S}{\partial x^j_*})[1+2\kappa(r-r_h)],\cr
Y&=&g^{jk}(\frac{\partial S}{\partial x^j_*})(\frac{\partial S}{\partial x^k_*})+\mu^2_0.
\end{eqnarray}
Multiplying by the factor $1/[r_{h}g^{01}\{1+2\kappa(r-r_{h}\}-rg^{0j}r_{h,j}]$ to both sides of Eq. (58) and assuming the resulting coefficient of $(\frac{\partial S}{\partial r_*})^2$ tends to unity near the event horizon, then we get
\begin{eqnarray}
\lim_{r\longrightarrow r_h} \frac{g^{11}r^{2}_{h}\{1+2\kappa(r-r_{h})\}^2-2g^{1j}rr_{h}r_{h,\phi}\{1+2\kappa(r-r_{h})\}}{2\kappa r_{h}(r-r_{h})[r_{h}g^{01}\{1+2\kappa(r-r_{h})\}-rg^{0j}r_{h,j}]}=1.
\end{eqnarray}
Eq. (60) is an indeterminate form of $0/0$ near the black hole event horizon and using L' Hospital's rule and adjusting parameter $\kappa$, the surface gravity obtained from relativistic Hamilton-Jacobi equation is given by
\begin{eqnarray}
\kappa=\frac{\frac{\partial g^{11}}{\partial r}-2\frac{\partial g^{1j}}{\partial r}r_{h,j}
+\frac{\partial g^{jk}}{\partial r}r_{h,j}r_{h,k}}{2[g^{01}-2g^{11}+(2g^{1j}-g^{0j})r_{h,j}]}
+\frac{g^{jk}r_{h,j}r_{h,k}-g^{1j}r_{h,j}}{r_{h}[g^{01}-2g^{11}+(2g^{1j}-g^{0j})r_{h,j}]}\cdot
\end{eqnarray}
which is equal to the surface gravity obtained from Klein-Gordon equation, Maxwell's equations and Dirac equations as given in Eqs. (10), (26) and (36).
Using the generalized tortoise coordinate transformation in Eq. (57) and defining $\frac{\partial \Phi}{\partial u_*}=-\omega, \frac{\partial \Phi}{\partial \theta_*}=K_{\theta}, \frac{\partial \Phi}{\partial \phi_*}=K_{\phi}$, for real $\frac{\partial \Phi}{\partial r_*}$ the distribution of energy levels of the particles is given by
\begin{eqnarray}
\omega\geq\omega^+\,\,\,\,\,\,\,\,\,\,\, {\rm and}\,\,\,\,\,\,\,\,\,\,\,\,\,\,\omega\leq\omega^{-}.
\end{eqnarray}
  Near the black hole event horizon there exist seas of positive and negative energy states and a forbidden energy gap. Penrose [56] proposed that a particle entering close to the surface of the event horizon decays into two particles - one particle having positive energy escapes to infinity and the other particle having negative energy gets absorbed by the black hole. A quantum analogue as spontaneous pair creation was proposed by Zel'dovich [57] in the Kerr black hole. The energy states must satisfy the condition $\omega^{-}\leq\omega\leq\omega^+$ at the forbidden region. The maximum value of the negative energy state after overlapping of the positive and negative energy states at the surface of the event horizon is
\begin{eqnarray}
\omega_h&=&K_{\theta}\frac{g^{12}-g^{2j}r_{h,j}}{g^{01}-g^{0j}r_{h,j}}+K_{\phi}\frac{g^{13}-g^{3j}r_{h,j}}{g^{01}-g^{0j}r_{h,j}}.
\end{eqnarray}
The width of the forbidden energy approaches to zero near the event horizon. This indicates that there exists a crossing of the positive and the negative energy levels near the event horizon [38]. When $\omega_{h}>\mu_0$, the particle can escape to infinity from the black hole event horizon. The Starobinskii-Unruh process (spontaneous radiation) must occur in the region near the black hole event horizon [57-61]. It indicates that the incident negative energy particle will become emerging positive energy particle via quantum tunneling effect. From this result, there is radiation near the event horizon. This type of radiation is independent of the temperature of the black hole and the type of quantum effect is a non-thermal.

 {\bf Application of this theory}. The line element describing general non-stationary symmetrical black hole in retarded time coordinates $(u, r, \theta, \phi)$ is defined by

\begin{eqnarray}
ds^2&=&g_{00}du^2+2g_{01}du dr+2g_{02}du d \theta+ g_{03}du d \phi + g_{13}dr d \phi\cr&&+ g_{22}d \theta^2+2 g_{23}d \theta d\phi+ g_{33}d\phi^2\cdot
\end{eqnarray}
Using Eqs. (4), (5) and (64), the expression of temperature in general non-stationary axial symmetrical black hole is
\begin{eqnarray}
T=\frac{1}{4\pi r_{h}}\Big[\frac{r_{h}(\frac{\partial g^{11}}{\partial r}-2\frac{\partial g^{1j}}{\partial r}r_{h,j}
+\frac{\partial g^{jk}}{\partial r}r_{h,j}r_{h,k})+2(g^{jk}r_{h,j}r_{h,k}-g^{1j}r_{h,j})}{\{g^{01}-2g^{11}+(2g^{1j}-g^{0j})r_{h,j}\}}\Big]
\end{eqnarray}
and the expression of chemical potential is defined by
\begin{eqnarray}
\omega_0&=&K_{\theta}\frac{g^{12}-g^{2j}r_{h,j}}{g^{01}-g^{0j}r_{h,j}}+K_{\phi}\frac{g^{13}-g^{3j}r_{h,j}}{g^{01}-g^{0j}r_{h,j}}\cdot
\end{eqnarray}
From Eqs. (63) and (66), we observe that the chemical potential derived from scalar particle is equal to the highest energy of the negative-energy state.

In particular, the line element of Kerr black hole in retarded time coordinate is given by [62]

\begin{eqnarray}
ds^2&=&(1-2MrR^{-2})du^2+2du dr +4arMR^{-2}\sin^2\theta du d\phi-2a\sin^2\theta dr d\phi\cr&&-R^2d\theta^2-\{(r^2+a^2)-\Delta a^2\sin^2\theta\}R^{-2}\theta^2 d\phi^{2},
\end{eqnarray}
where $\Delta=r^2-2Mr+a^2$ and $R^2=r^2+a^2\cos^{2}\theta$. The Kerr black hole has singularity at $\Delta=0$. The roots $r_h=M+\sqrt{M^2-a^2}$ and $r_h=M-\sqrt{M^2-a^2}$ represent external event horizon and internal Cauchy horizon respectively. From Eqs. (65) and (67), the Hawking temperature of non-stationary Kerr black hole is obtained as
\begin{eqnarray}
T&=&\frac{1}{2\pi}\frac{r^2_{h}(1+2r_{h,u})-r_hM(u_{0})-\{\Delta_h+r_{h,u}(r^2_h+a^2)+ar_{h,\phi}\}}{r_h\{(r^2_{h}+a^2+a^2\sin^2\theta_{0})(1+2r_{h,u})+Z\}}
\end{eqnarray}
where $Z=2r^2_{h,\theta}+r_{h,\phi}(4ar_{h,u}+\frac{2r_{h,\phi}}{\sin^2\theta_{0}}+3a)$
and the chemical potential is
\begin{eqnarray}
\omega_0&=&\frac{K_{\theta}r_{h,\theta}}{[(r^2_h+a^2)+a^2\sin^2\theta_{0}r_{h,u}+ar_{h,\phi}]}\cr&&+\frac{K_{\phi}(r_{h,u}+a^2\sin^2\theta_{0}+r_{h,\phi})}{\sin^2\theta_{0}[(r^2_h+a^2)+a^2\sin^2\theta_{0}r_{h,u}+ar_{h,\phi}]}\cdot
\end{eqnarray}
This indicates that the chemical potential derived from scalar particles is equal to highest energy of the negative-energy state. This was mentioned at the beginning and has been shown by direct calculation to a special case.

8. {\bf Discussion}.
 The total interaction energy of scalar particles of Klein-Gordon, Maxwell's electromagnetic field equations and Dirac particles in general non-stationary black hole is given by
\begin{eqnarray}
\omega_0&=&K_{\theta}\frac{g^{12}-g^{2j}r_{h,j}}{g^{01}-g^{0j}r_{h,j}}+K_{\phi}\frac{g^{13}-g^{3j}r_{h,j}}{g^{01}-g^{0j}r_{h,j}}+L_4
\end{eqnarray}
where
\begin{eqnarray}
&&L_4=0
\end{eqnarray}
for $\Psi=\Phi$ (Klein-Gordon scalar particle),
\begin{eqnarray}
L_4&=& \frac{-i}{(g^{01}-g^{0j}r_{h,j})}[\ell^\nu(n^1_{,\nu}-n^j_{,\nu}r_{h,j})-m^\nu(\ell^1_{,\nu}-\ell^j_{,\nu}r_{h,j})
-2\rho(n^1-n^jr_{h,j})\cr&&+(\mu-2\gamma)(\ell^1_{,\nu}-\ell^j_{,\nu}r_{h,j})-(\pi-2\alpha)(m^1-m^jr_{h,j})
+2\tau(\bar{m}^1-\bar{m}^jr_{h,j})\cr&&
+\frac{(\bar{m}^1-\bar{m}^jr_{h,j})}{(\ell^1-\ell^jr_{h,j})}\{m^\nu(\ell^1_{,\nu}-\ell^j_{,\nu}r_{h,j})
-\ell^\nu(\bar{m}^1_{,\nu}-\bar{m}^jr_{h,j})\}\cr&&
-\{\sigma(\bar{m}^1-\bar{m}^jr_{h,j})
-\kappa(n^1-n^jr_{h,j})\}\frac{(n^1-n^jr_{h,j})(\bar{m}^1-\bar{m}^jr_{h,j})}{(m^1-m^jr_{h,j})(\ell^1-\ell^jr_{h,j})}]
\end{eqnarray}
for $\Psi=\phi_0$ (Maxwell's electromagnetic field),
\begin{eqnarray}
L_4&=&D_4=\frac{-i}{(g^{01}-g^{0j}r_{h,j})}[-2\rho(n^1-n^jr_{h,j})+(\mu-2\gamma)(\ell^1-\ell^jr_{h,j})
+2\tau(\bar{m}^1-\bar{m}^jr_{h,j})\cr&&-(\pi-2\alpha)(m^1-m^jr_{h,j})
+n^\nu(\ell^1_{,\nu}-\ell^j_{,\nu}r_{h,j})-\bar{m}^\nu(m^1_{,\nu}-m^j_{\nu}r_{h,j})\cr&&
+\frac{(\ell^1-\ell^jr_{h,j})}{(\bar{m}^1-\bar{m}^jr_{h,j})}\{\bar{m}^\nu(n^1_{,\nu}-n^j_{,\nu}r_{h,j})
-n^\nu(\bar{m}^1_{,\nu}-\bar{m}^j_{,\nu}r_{h,j})\}\cr&&
+\{-\sigma(\bar{m}^1-\bar{m}^jr_{h,j})
+\kappa(n^1-n^jr_{h,j})\}\frac{(n^1-n^jr_{h,j})}{(m^1-m^jr_{h,j})}]
\end{eqnarray}
for $\Psi=\phi_1$ and
\begin{eqnarray}
L_4&=&D_4=\frac{-i}{(g^{01}-g^{0j}r_{h,j})}[n^\nu(\ell^1_{,\nu}-\ell^j_{,\nu}r_{h,j})-\bar{m}^\nu(m^1_{,\nu}-m^j_{,\nu}r_{h,j})
\cr&&-2\pi(m^1_{,\nu}-m^j_{,\nu}r_{h,j})+(\tau-2\beta)(\bar{m}^1_{,\nu}-2(\rho-2\epsilon)(n^1_{,\nu}-n^j_{,\nu}r_{h,j})
\cr&&+2\mu(\ell^1_{,\nu}-\ell^j_{,\nu}r_{h,j})
+\frac{(\ell^1-\ell^jr_{h,j})} {(\bar{m}^1-\bar{m}^jr_{h,j})}\frac{(m^1-m^jr_{h,j})}{(n^1-n^jr_{h,j})}\{\lambda(n^1_{,\nu}-n^j_{,\nu}r_{h,j})\cr&&
-\nu(\bar{m}^1_{,\nu}-\bar{m}^j_{,\nu}r_{h,j})\}-2\{n^\nu(\bar{m}^1-\bar{m}^jr_{h,j})
\cr&&-\bar{m}^\nu(n^1_{,\nu}-n^j_{,\nu}r_{h,j})\}\frac{(m^1-m^jr_{h,j})}{(n^1-n^jr_{h,j})}]
\end{eqnarray}
for $\Psi=\phi_2$.
 \begin{eqnarray}
L_4&=&\frac{-i}{(g^{01}-g^{0j}r_{h,j})}[\ell^\nu(n^1_{,\nu}-n^j_{,\nu}r_{h,j})-m^\nu(\ell^1_{,\nu}-\ell^j_{,\nu}r_{h,j})
-2\rho(n^1-n^jr_{h,j})\cr&&+(\mu-2\gamma)(\ell^1_{,\nu}-\ell^j_{,\nu}r_{h,j})-(\pi-2\alpha)(m^1-m^jr_{h,j})
+2\tau(\bar{m}^1-\bar{m}^jr_{h,j})\cr&&
+\frac{(\bar{m}^1-\bar{m}^jr_{h,j})}{(\ell^1-\ell^jr_{h,j})}\{m^\nu(\ell^1_{,\nu}-\ell^j_{,\nu}r_{h,j})
-\ell^\nu(\bar{m}^1_{,\nu}-\bar{m}^jr_{h,j})\}\cr&&
-\{\sigma(\bar{m}^1-\bar{m}^jr_{h,j})
-\kappa(n^1-n^jr_{h,j})\}\frac{(n^1-n^jr_{h,j})(\bar{m}^1-\bar{m}^jr_{h,j})}{(m^1-m^jr_{h,j})(\ell^1-\ell^jr_{h,j})}]
\end{eqnarray}
for $\Psi=f_1$ ( Dirac Particles)
\begin{eqnarray}
L_4&=&\frac{-i}{(g^{01}-g^{0j}r_{h,j})}[(\mu-\gamma)(\ell^1-\ell^jr_{h,j})-(\pi-\alpha)(m^1-m^jr_{h,j})
\cr&&+\ell^\nu(n^1_{,\nu}-n^j_{,\nu})
+(\bar{\epsilon}-\bar{\rho})(n^1-n^jr_{h,j})+\ell^\nu(n^1_{,\nu}-n^j_{,\nu})-m^\nu(\bar{m}^1-\bar{m}^j_{,\nu})
\cr&&-(\bar{\epsilon}-\bar{\rho}-\epsilon+\rho)(n^1_{,\nu}-n^j_{,\nu})
-\ell^\nu\frac{(m^1_{,\nu}-m^j_{,\nu})}{(\bar{m}^1-\bar{m}^j_{h,j})}(n^1-n^jr_{h,j})\cr&&+\{m^\nu
\frac{(\ell^1_{,\nu}-\ell^j_{,\nu})}{(m^1-m^jr_{h,j})}-(\beta-\tau-\bar{\pi}
+\bar{\alpha})\frac{(\ell^1-\ell^jr_{h,j})}{(m^1-m^jr_{h,j})}\}(n^1-n^jr_{h,j})]
\end{eqnarray}
for $\Psi=f_2$

 This chemical potential $\omega_0$ is composed of two parts. The sum of the first two terms i.e, $2K_{\theta}\frac{g^{12}-g^{2j}r_{h,j}}{g^{01}-g^{0j}r_{h,j}}+2K_{\phi}\frac{g^{13}-g^{3j}r_{h,j}}{g^{01}-g^{0j}r_{h,j}}$ is the rotational energy arising from the coupling between different components of generalized momentum of Maxwell's electromagnetic field or Dirac particles and different rotations of black hole. The second term $L_4$ gives a new extra spin-rotation coupling and spin acceleration coupling effect. The physical origin of extra coupling effect comes from the interaction between intrinsic spin of particles and generalized momentum of evaporating black hole but it has no classical correspondence [29, 52]. The value of $L_4$ will vanish for the stationary black hole and scalar particle. When $L_4=0$, Eqs. (63) and (70) show the chemical potential derived from the thermal radiation spectrum of Maxwell's electromagnetic field or Dirac particle is equal to highest energy of the negative energy state of scalar particles in general non-stationary black hole. This brings out a clear relationship between the two kinds of radiation processes of black holes.

On the other hand, the quantitative causal relation with no-loss-no-gain character would be satisfied by a lot of general physical process [63, 64]. Using the no-loss-no-gain homeomorphic map transformation satisfying causal relation, Ref. [65] derived the exact strain tensor formulas in Weitzenbock manifold. Ref. [66] investigated the cosmic quantum birth by studying the Wheeler-DeWitt equation  which satisfies quantitative causal relation. In fact, some effect of change (cause) of some quantities in (2) must result in the relative changes (cause) in (2) so as to keep no-loss-no-gain in the right hand side of (2), that is zero. Simarly Eqs.(6), (13), (15-17), (27-29) and (57) must satisfy the quantitative causal relation in the same way. Hence the findings in this paper are consistent.

Refs. [29-31] proposed the tortoise coordinate transformation which is applicable to the black hole event horizon
 \begin{eqnarray}
r_*&=&r+\frac{1}{2\hat{\kappa}(u_0,\theta_0, \phi_0)}ln\Big\{r-r_h(u,\theta,\phi)\Big\},\cr
u_*&=&u-u_0,\,\,\,\,\,\, \theta_*=\theta-\theta_0,\,\,\,\,\,\,\,\,\,\,\,\,\,\,\phi_*=\phi-\phi_0,
\end{eqnarray}
it gives the surface gravity at the event horizon as
\begin{eqnarray}
\hat{\kappa}(u_0, \theta_0, \phi_0)&=&\frac{\frac{\partial g^{11}}{\partial r}-2\frac{\partial g^{1j}}{\partial r}r_{h,j}
+\frac{\partial g^{jk}}{\partial r}r_{h,j}r_{h,k}}{2[g^{01}-2g^{11}+(2g^{1j}-g^{0j})r_{h,j}]},
\end{eqnarray}
which is different from one given by Eqs. (10), (26) and (36) corresponding to the generalized tortoise coordinate transformation (3). Using the different tortoise coordinate transformations (3) and (77), it has been observed that the surface gravities of the general non-stationary black hole for Klein-Gordon scalar particles, Maxwell's electromagnetic field equations, Dirac equations and relativistic Hamilton-Jacobi equation near the black hole event horizon are the same as shown in (10), (26), (36) and (61) under the transformation (3) but they are different for different tortoise coordinate transformations.
From Eqs. (61) and (78), the surface gravity can be written as
\begin{eqnarray}
\kappa(u_0, \theta_0, \phi_0)=\hat{\kappa}(u_0, \theta_0, \phi_0)+\xi(u_0, \theta_0, \phi_0),
\end{eqnarray}
where $\kappa(u_0, \theta_0, \phi_0)$ and $\hat{\kappa}(u_0, \theta_0, \phi_0)$ are the surface gravities obtained from tortoise coordinate transformations (3) and (77). Similarly, Eq. (52)  can be expressed as
\begin{eqnarray}
T(u_0, \theta_0, \phi_0)&=&\hat{T}(u_0, \theta_0, \phi_0)+\frac{1}{2\pi}\xi(u_0, \theta_0, \phi_0)
\end{eqnarray}
where $\hat{T}(u_0, \theta_0, \phi_0)$ is the thermal radiation temperature under tortoise coordinate transformation (77) and   $\xi(u_0, \theta_0, \phi_0)$ denotes the difference of thermal radiation under the different tortoise coordinate transformations and it is given by
\begin{eqnarray}
\xi(u_0, \theta_0, \phi_0)=\frac{g^{jk}r_{h,j}r_{h,k}-g^{1j}r_{h,j}}{r_{h}[g^{01}-2g^{11}+(2g^{1j}-g^{0j})r_{h,j}]}.
\end{eqnarray}
 According to Ref. [43, 46], the correction rate under the different tortoise coordinate transformations is given by
\begin{eqnarray}
\Upsilon(u_0, \theta_0, \phi_0)=\frac{2(g^{jk}r_{h,j}r_{h,k}-g^{1j}r_{h,j})}{r_h(\frac{\partial g^{11}}{\partial r}-2\frac{\partial g^{1j}}{\partial r}r_{h,j}
+\frac{\partial g^{jk}}{\partial r}r_{h,j}r_{h,k})}.
\end{eqnarray}
If $\xi(u_0, \theta_0, \phi_0)$ approaches to zero, the two surface gravities are equal due to the tortoise coordinate transformations (3) and (77). This indicates that different tortoise coordinate transformations correspond to different Hawking radiation temperatures in a non-stationary rotating black hole space time [45].
 Eq. (80) implies that when $\hat{T}(u_0, \theta_0, \phi_0)$ become zero, the temperature $T(u_0, \theta_0, \phi_0)$ will not be vanished due to the presence of extra term $\xi(u_0, \theta_0, \phi_0)$. From this research paper, we conclude that the generalized tortoise coordinate transformation (3) provides an alternative and convenient way to obtain Hawking radiation and is also generalization and development of the works of [36, 52] about quantum radiation for general non-stationary black holes.

9. {\bf Conclusions}. \setcounter {equation}{0}
\renewcommand {\theequation}{7.\arabic{equation}}
In this paper the thermal radiation and chemical potential is investigated using the generalized tortoise coordinate transformation in general non-stationary black hole. The locations of the horizon and the thermal radiation near the black hole event horizon are determined. It has been found that they are functions of retarded time co-ordinate $u$ and different angles $\theta, \phi$. By adjusting properly the value of $\kappa$, the second order form of Klein-Gordon equation, the three second order form of Maxwell's electromagnetic field equations and the two second order form of Dirac equations are transformed into a standard form of wave equation near the event horizon $r=r_h$. Separating the variables of the wave equation and applying the well known Damour-Ruffini-Sannan method, we determine accurately the thermal radiation and chemical potential at the event horizon $r=r_h$ in general non-stationary black hole. It is also observed that the constant term $L_4$ appeared in the expression of chemical potential which gives the interaction between the intrinsic spin of the particles and the generalized momentum in general non-stationary black hole but it is found to be absent in the chemical potential derived from Klein-Gordon scalar particles. This paper reveals the chemical potential obtained from the thermal radiation spectrum of Klein-Gordon scalar particles is equal to the highest energy of the negative energy states of the non-thermal radiation in general non-stationary black hole near the event horizon. Besides, under the generalized coordinate transformation a constant term $\xi(u_0, \theta_0, \phi_0)$ is appeared in the expression of surface gravity and the thermal radiation of black hole near the event horizon. As $\xi(u_0, \theta_0, \phi_0)$ tends to zero, our results are consistent with already results obtained by Hua and Huang [36, 52]. In conclusion, the generalized tortoise coordinate transformation is found to be more reliable and accurate in the study of thermal radiation spectrum in general non-stationary black hole near the event horizon $r=r_h$.

{\bf Conflict of Interests}

The author declares that there is no conflict of interests regarding the publication of this paper.

{\bf Acknowledgements}: The author would like to thank Prof. Ng. Ibohal, Prof. N. Nimai and Dr I. Ablu for illuminating discussions.

\end{document}